\documentclass[journal=jpccck,manuscript=article]{achemso}

\usepackage{amsmath}
\usepackage{subcaption}
\usepackage{hyperref} 

\title{AI-Assisted Physics-Informed Predictions of Degradation Behavior of Polymeric Anion Exchange Membranes}

\author{William Schertzer}
\affiliation{Georgia Institute of Technology, School of Materials Science and Engineering}
\author{Mohamed Al Otmi}
\affiliation{University of Florida, Department of Chemical Engineering}
\author{Janani Sampath}
\affiliation{University of Florida, Department of Chemical Engineering}
\author{Ryan P. Lively}
\affiliation{Georgia Institute of Technology, School of Chemical and Biomolecular Engineering}
\author{Rampi Ramprasad}
\affiliation{Georgia Institute of Technology, School of Materials Science and Engineering}
\email{rampi.ramprasad@mse.gatech.edu}


\begin{document}

\begin{abstract}

The global transition to hydrogen-based energy infrastructures faces significant hurdles. Chief among these are the high costs and sustainability issues associated with acid-based proton exchange membrane fuel cells. Anion exchange membrane (AEM) fuel cells offer promising cost-effective alternatives, yet their widespread adoption is limited by rapid degradation in alkaline environments. Here, we develop a framework that integrates mechanistic insights with machine learning, enabling the identification of generalized degradation behavior across diverse polymeric AEM chemistries and operating conditions. Our model successfully predicts long-term hydroxide conductivity degradation (up to 10,000 hours) from minimal early-time experimental data. This capability significantly reduces experimental burdens and may expedite the design of high-performance, durable AEM materials.

\end{abstract}

\flushbottom
\maketitle
\thispagestyle{empty}

\section*{Introduction}
Increasing global demand for clean energy has spurred widespread interest in the development of cost-effective and efficient fuel cell technology \cite{dekel_review_2018, tran_design_2024}. Although proton exchange membrane (PEM) fuel cells have received the majority of research attention to date, their reliance on costly and environmentally persistent perfluorinated polymers (e.g., Nafion) and platinum-based catalysts significantly limits their scalability and sustainability \cite{dekel_review_2018, noauthor_technical_nodate}.

Anion exchange membrane (AEM) fuel cells operate under alkaline conditions and conduct hydroxide ions with faster reaction kinetics compared to their PEM counterparts, enabling the use of inexpensive, fluorine-free hydrocarbon-based polymers and non-precious metal catalysts \cite{lee_robust_2015,hossen_state---art_2023}. These potential cost advantages have driven significant research interest in AEMs over the past two decades, positioning them as promising alternatives to conventional acid-based fuel cells \cite{hossen_state---art_2023,merle_anion_2011, arges_anion_2018,mandal_recent_2021}. This shift not only reduces cost and environmental burden but also opens pathways for more sustainable polymer design \cite{mandal_recent_2021}. However, the commercialization of AEM fuel cells is still hindered by major challenges, notably the chemical and mechanical instability of AEMs in alkaline media, which leads to rapid degradation and failure far before the 20{,}000--25{,}000-hour target operational lifetime set by the U.S. Department of Energy \cite{noauthor_hydrogen_2024}. This challenge is coupled with the need for membranes with high hydroxide conductivity, e.g., greater than 100 mS/cm \cite{lee_robust_2015}. Unfortunately, chemical durability and ionic conductivity are material properties that are often in conflict \cite{willdorf-cohen_alkaline_2023}.

A significant body of research has focused on improving individual aspects of AEM performance\textemdash such as enhancing ion exchange capacity (IEC), reducing water uptake (WU) and swelling ratio (SR), and improving hydroxide conductivity \cite{dekel_review_2018,mandal_recent_2021, raut_effect_2023}. Our previous work contributed to this area by leveraging machine learning and atomistic models learning to predict these static properties and identify fluorine-free AEM candidates that strike a balance between performance and stability \cite{al_otmi_investigating_2024, schertzer_ai-driven_2025, tran_informatics-driven_2023}. Yet, while such models capture initial performance, they offer limited insight into the long-term degradation behavior that ultimately governs membrane viability in real-world applications, and they fail to capture the interplay between these individual aspects as they relate to long-term durability \cite{mustain_durability_2020, jiang_durable_2023, song_upscaled_2023}.

Recent advances in machine learning research have led to the development of various types of robust algorithms suitable for many science and engineering applications \cite{batra_emerging_2021, nistane_polymer_2025}. In prior work, we applied Gaussian Process Regression (GPR) to assess the extrapolative capability of ML models for AEMs and to quantify the chemical diversity of the available dataset \cite{schertzer_ai-driven_2025}. That analysis revealed that the limited diversity of reported chemistries imposes inherent constraints on the accuracy of predictions for entirely novel formulations. Building on this insight, the present study shifts focus toward uncovering trends within the existing chemical space using a Physics-Enforced Neural Network (PENN).

In this study, we extend our informatics-based approach to address a critical missing dimension in AEM research: time-dependent degradation. Specifically, we focus on the evolution of hydroxide conductivity under prolonged alkaline exposure, a key indicator of chemical and structural breakdown in AEMs \cite{willdorf-cohen_alkaline_2023, mustain_durability_2020}. Although prior studies have investigated the degradation of specific AEMs by introducing structural modifications (e.g., flexible spacers, crosslinking, branching, or inorganic additives \cite{dekel_review_2018, lee_robust_2015, hossen_state---art_2023, mandal_recent_2021}), these efforts have largely remained fragmented. They often focus on narrow design variations and isolated degradation mechanisms, making it difficult to generalize findings across the broader AEM landscape.

\begin{figure}[t]
\centering
\includegraphics[width=1.0\linewidth]{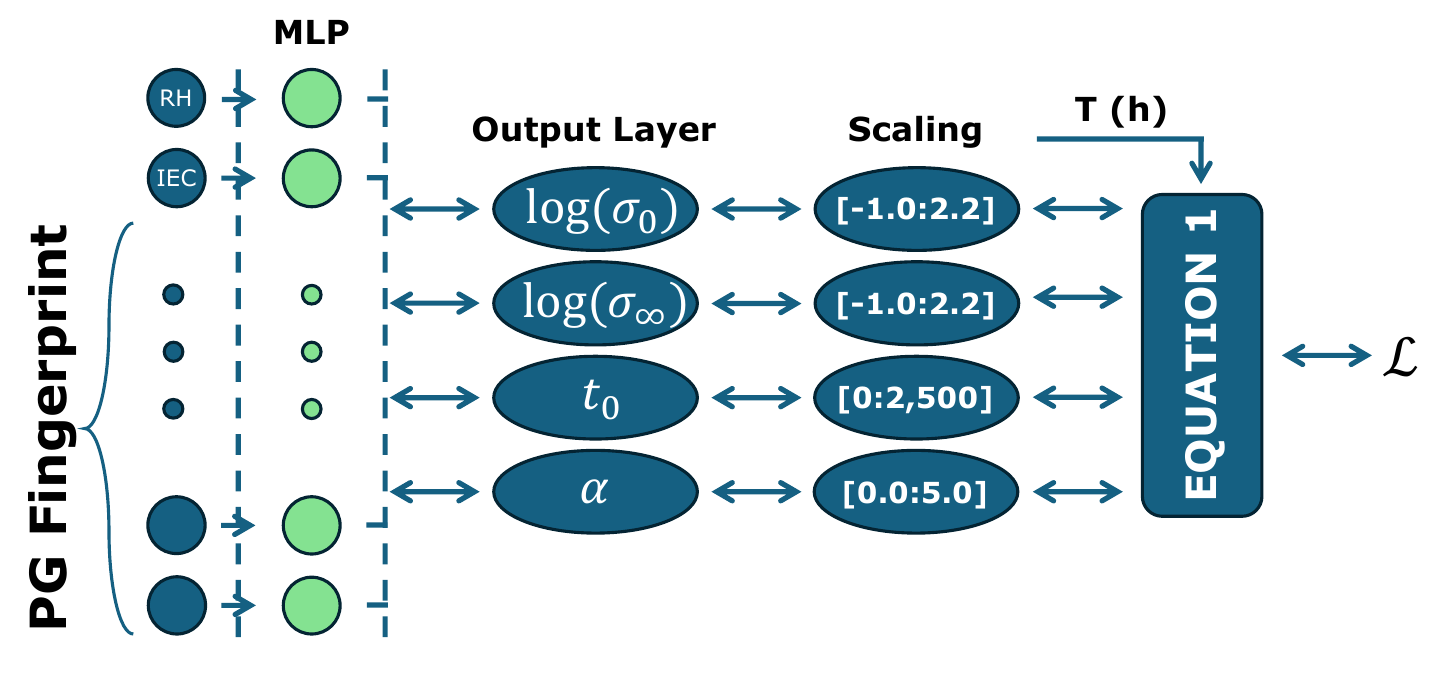}
\caption{Schematic of the PENN architecture. Polymer genome fingerprints and environmental features are input to a multilayer perceptron (MLP), which predicts four physically meaningful degradation parameters: initial conductivity \(\sigma_0\), limiting conductivity \(\sigma_{\infty}\), characteristic time \(t_0\), and decay shape parameter \(\alpha\). These parameters are then passed through a mechanistic degradation equation and compared to experimental time series to guide training via a physics-informed loss function.}
\label{fig:PENN_architecture}
\end{figure}

To address this gap, we have created a comprehensive database of time-resolved hydroxide conductivity measurements from the literature. The database encompasses a wide diversity of polymer backbones, cationic groups, solvents, additives, temperatures, and relative humidities. Upon observing time-dependent hydroxide conductivity trends across a breadth of chemistries and environmental conditions, we identified a consistent empirical relationship for the degradation of hydroxide conductivity in any AEM system exposed to alkaline media for prolonged periods. Equation \ref{deg} is proposed to describe the degradation of the hydroxide conductivity of AEMs: 

\begin{equation}\label{deg}
\log\sigma(t) = \log\sigma_{\infty} + \frac{\log\sigma_0 - \log\sigma_\infty}{1 + \left(\frac{t}{t_0}\right)^{\alpha}}
\end{equation}

In this equation, \(\sigma_0\) represents the initial hydroxide conductivity at \(time=0\), while \(\sigma_{\infty}\) is the limiting conductivity at long times or under equilibrium conditions. The parameter \(t_0\) is a characteristic time scale that governs the halfway drop-off point from \(\log\sigma_0\) to \(\log\sigma_{\infty}\), and \(\alpha\) is a shape parameter that determines the steepness of the decay curve. This equation captures the time-dependent degradation of conductivity due to complex chemical processes under alkaline conditions, reflecting the initial performance, long-term stability, and rate of performance decay of AEMs.

We then introduce a PENN framework designed to uncover universal degradation trends from this heterogeneous dataset by predicting the four parameters (\(\sigma_0, \sigma_{\infty}, t_0, \alpha\)) for each AEM sample. Figure \ref{fig:PENN_architecture} depicts the PENN architecture employed here: Polymer Genome fingerprints \cite{kim_polymer_2018} are concatenated with environmental variables and passed through a multilayer perceptron (MLP) to predict the four output parameters of Equation \ref{deg}. These predicted parameters are scaled to physically reasonable ranges and then passed to the loss function along with the measurement time and ground truth conductivity value for each sample to achieve model training.

AEM degradation is driven by complex coupled chemical processes\textemdash{}such as \(\beta\)-elimination, nucleophilic substitution, and polymer backbone scission\textemdash{}that occur under alkaline conditions and are influenced by the chemistry of the polymer, the IEC, the degree of crosslinking, the presence of stabilizing or destabilizing additives, and the operating environment (e.g., temperature and relative humidity) \cite{dekel_review_2018, willdorf-cohen_alkaline_2023}. These intertwined effects make it difficult to isolate causal relationships using traditional empirical studies and underscore the need for a unified framework capable of modeling long-term behavior across a chemically diverse set of AEMs \cite{hossen_state---art_2023}.

If Equation \ref{deg} is true, then appropriately normalizing the degradation curves would reveal a universal behavior across multiple samples. Equation \ref{deg_normalized} shows the normalized degradation equation, in which the degradation curves of all systems may collapse onto a single master curve, suggesting a universal degradation behavior that transcends specific chemistries and environmental conditions:

\begin{equation}\label{deg_normalized}
\log\hat\sigma(t) = \frac{1}{1 + \hat{t}}, \
\text{where } \log\hat{\sigma} = \frac{\log\sigma - \log\sigma_{\infty}}{\log\sigma_0 - \log\sigma_{\infty}}, \
\text{and } \hat{t} = \left[\frac{t}{t_0}\right]^{\alpha}
\end{equation}

By passing the predicted parameters and measurement time for each sample through Equation \ref{deg_normalized} and visually comparing the \(\hat{\sigma}\) vs \(\hat{t}\) curves, we show that there is indeed a predictable set of parameters for each sample such that the normalized predicted and observed hydroxide conductivity are in agreement across the observed range of AEM formulations.

Going further towards helping make efficient engineering decisions, we demonstrate the ability of our PENN framework to distinguish between different degradation modes and to extrapolate long-term degradation behavior from short-term data. A comparison of PENN vs baseline NN and GPR models (in which Equation \ref{deg} is not enforced) shows that our proposed method excels at identifying samples that exhibit bimodal degradation patterns (rapid initial degradation followed by smooth, gradual degradation), where NN overfits (predicting nonphysical increasing trends) and GPR oversimplifies (predicting nonphysical smooth degradation trends). We achieve accurate predictions of hydroxide conductivity at several thousands of hours using only a few hundred hours of early-time measurements, and we show the jump in forecasting predictions when using PENN compared to NN and GPR. These capabilities have the potential to drastically reduce the experimental burden associated with long-term stability testing.

\section*{Methods and Materials}
\subsection*{Dataset}

The training dataset used in this study contains over 5,200 data points manually extracted from tables and figures in academic articles with the help of the WebPlotDigitizer tool \cite{noauthor_automerisio_nodate}. The dataset, along with the associated DOI of each entry, is publicly available on the \href{https://github.com/Ramprasad-Group/polyVERSE}{polyVERSE GitHub}. Both static and time-resolved property measurements are recorded, with over 2,200 unique hydroxide conductivity data points, each corresponding to a distinct AEM system. The property values were recorded at time points spanning from initial synthesis and preparation (t = 0 h) to complete membrane failure (t \(\geq\) 10,000 h) under various experimental conditions. The dataset contains 112 unique profiles of time-resolved hydroxide conductivity measurements of AEM formulations. In many of these profiles, we observed an initial rapid increase in hydroxide conductivity prior to its eventual decay. This "waking up" effect is likely due to membrane hydration. To ensure consistent model training and reflect the true onset of degradation, we shifted the time axis such that \(t=0\) corresponds to the point of maximum conductivity for each sample. A summary of the number of data points available for each property, along with the percentage of each property in the dataset and the percentage of time-dependent data, is provided in Table \ref{tab:dataset_stats}. 

In preparation for machine learning analysis, each profile was annotated with its unique combination of chemical and environmental descriptors, reflecting both polymer composition and test conditions. These include:
\begin{enumerate}
    \item Monomer structure: Represented as SMILES strings (SMILES1–SMILES3) for each monomeric repeat unit in the copolymer.
    \item Monomer composition: Mole fractions (c1–c3) indicating the proportion of each monomer in the statistical copolymer backbone.
    \item Theoretical ion exchange capacity (IEC): The number of active ion exchange sites per polymer repeat unit (reported in meq/g), calculated from polymer structure and composition, as described in our previous contribution \cite{schertzer_ai-driven_2025}.
    \item Relative humidity (RH): The ambient humidity (reported in percent) during conductivity measurement.
    \item Stability test temperature: The temperature (reported in \textdegree C) at which the degradation experiment was conducted.
    \item Measurement temperature: The temperature (reported in \textdegree C) at which the conductivity measurement was recorded.
    \item Solvent type and concentration: The identity of the solvent used during degradation testing (e.g., KOH, NaOH) and its reported concentration (reported as molarity [mol/L]).
    \item Additive type and concentration: The identity of the additive(s) used during degradation testing (e.g., stabilizers, crosslinkers, inorganic fillers) and the corresponding concentration (reported in wt \%).
    \item Time: The amount of time (reported in hours) that the sample has been submerged in a particular solvent.
\end{enumerate}

In addition to conductivity, the dataset also includes both static and time-resolved measurements for other key AEM properties (Ion Exchange Capacity, Water Uptake, Swelling Ratio, Tensile Strength, Elongation at Break, and Young's modulus), all of which have been discussed in our previous contribution. Although this work focuses exclusively on modeling the time evolution of hydroxide conductivity, these additional properties offer valuable insight into mechanical and transport degradation behavior. In the future, a multi-task framework that jointly models the degradation of conductivity, swelling, and mechanical performance will be explored to enable holistic lifetime prediction of AEMs given sparse time-resolved property data.

\begin{table}[ht]
\centering
\small  
\caption{Summary of dataset properties, including the percentage of dataset for each property, the percentage of that property with time-dependent data, and the number of data points for each property.}
\begin{tabular}{cccccccccc}
\hline
\textbf{Property} & \textbf{Number of Data Points} & \textbf{\% of Dataset} & \textbf{\% Time Dependent} \\
\hline
OH$^{-}$ Cond. (mS/cm) & 2,229 & 40.54 & 50.51  \\
Ion Exchange Capacity (meq/g)            & 1,485 & 27.53 & 30.58 \\
Water Uptake (wt\%)              & 627 & 11.40 & 3.03 \\
Swelling Ratio (\%)          & 521 & 9.53  & 2.20  \\
Tensile Strength (MPa)               & 171 & 3.11  & 10.52 \\
Elongation at Break (\%) & 163 & 2.96  & 8.30 \\
Young's Modulus (MPa)               & 73 & 1.32  & 0 \\
Total                  & 5,269 & 100 & 33.23 \\
\hline
\end{tabular}
\label{tab:dataset_stats}
\end{table}

\subsection*{Feature Engineering}
Chemical features were extracted from each sample using the Co-Polymer Genome fingerprinting scheme, which encodes the hierarchical structure of copolymers and has been shown to accurately predict a wide range of polymer properties, including hydroxide conductivity, water uptake, and swelling ratio \cite{kuenneth_copolymer_2021, kim_polymer_2018}. Each monomeric repeat unit is converted into a chemically informed descriptor vector, capturing atomic, structural, and electronic features. These vectors are combined into a single polymer fingerprint via a composition-weighted linear combination, reflecting the molar ratios of monomers in the copolymer backbone.

To this base representation, we append experimentally relevant environmental descriptors, including ion exchange capacity (IEC), relative humidity (RH), and temperature, as described in Equations (1)–(3) of our prior work \cite{schertzer_ai-driven_2025}. In the present study focused on modeling time-resolved degradation, we extend this fingerprinting framework to include new physicochemical and environmental features that influence degradation under alkaline conditions. These include:
\begin{enumerate}
    \item Stability test temperature, reflecting the thermal environment during degradation experiments.
    \item Solvent concentration, capturing the identity and strength of the alkaline medium (e.g., KOH, NaOH).
    \item Additive concentration, encompassing stabilizers, crosslinkers, or inorganic fillers added to enhance stability.
\end{enumerate}
Continuous-valued concentration features of each solvent and additive are used to encode the presence and relative amount of each component in the sample. The final feature vector, comprising polymer structure, environmental descriptors, and processing conditions, is fully normalized on a scale of [0:1] to ensure numerical stability and effective training.

\begin{figure}[!th]
    \centering
    \begin{subfigure}[t]{0.33\linewidth}
        \centering
        \includegraphics[width=\linewidth]{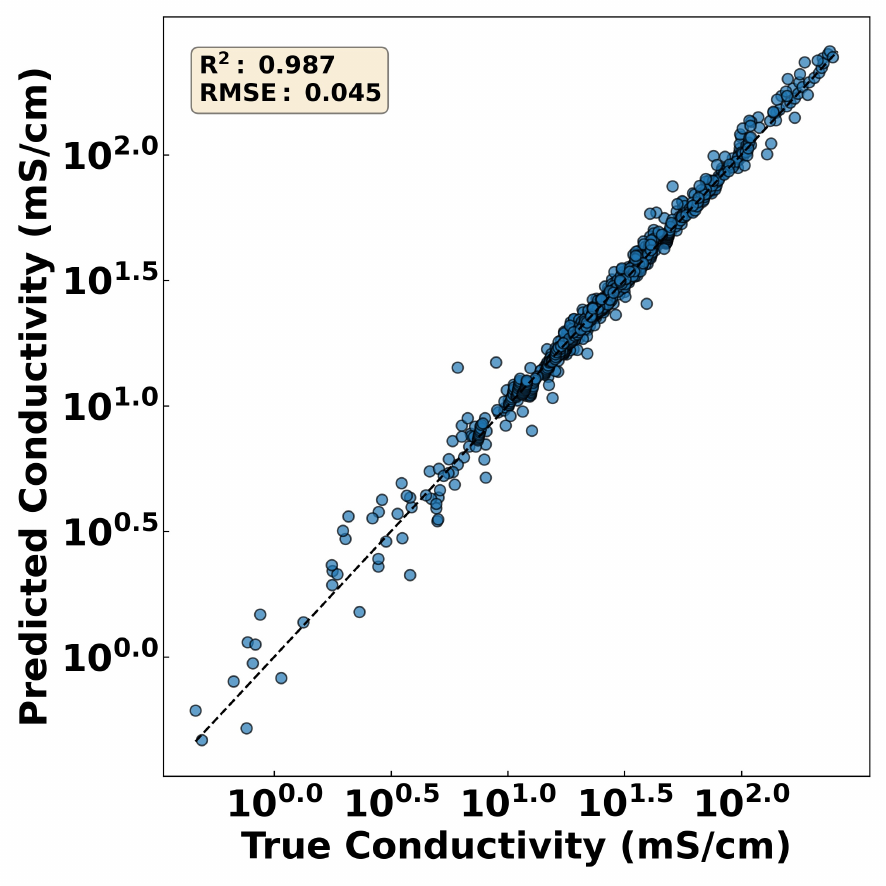}
        \caption{PENN parity plot}
        \label{fig:all_data_parity_penn}
    \end{subfigure}%
    \hfill
    \begin{subfigure}[t]{0.33\linewidth}
        \centering
        \includegraphics[width=\linewidth]{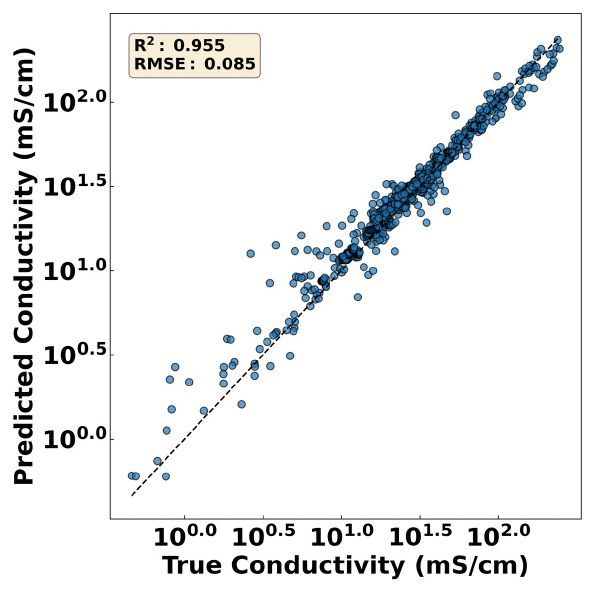}
        \caption{NN parity plot}
        \label{fig:all_data_parity_nn}
    \end{subfigure}%
    \hfill
    \begin{subfigure}[t]{0.33\linewidth}
        \centering
        \includegraphics[width=\linewidth]{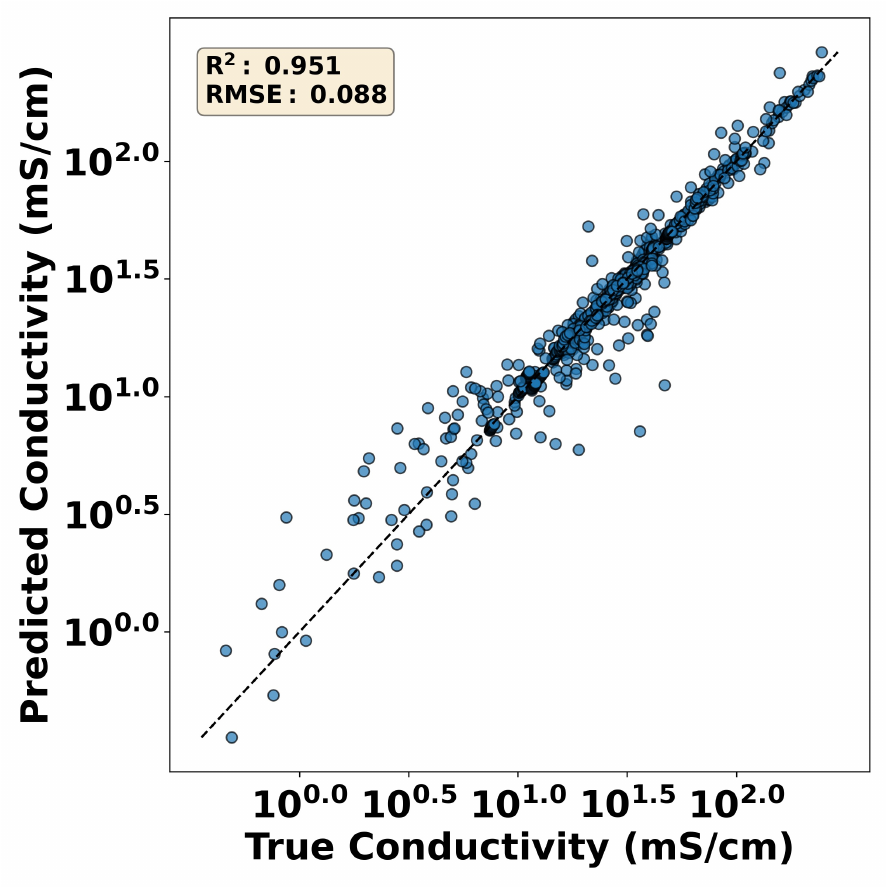}
        \caption{GPR parity plot}
        \label{fig:all_data_parity_gpr}
    \end{subfigure}
    
    \caption{Parity plots comparing predicted versus true hydroxide conductivity across all test samples using (a) PENN, (b) NN and (c) GPR models. All models show good accuracy and consistency across the range of predicted conductivity values when using the entire dataset for training.}
    \label{fig:parity_plots}
\end{figure}

\begin{figure}[!t]
    \centering
    \includegraphics[width=1\linewidth]{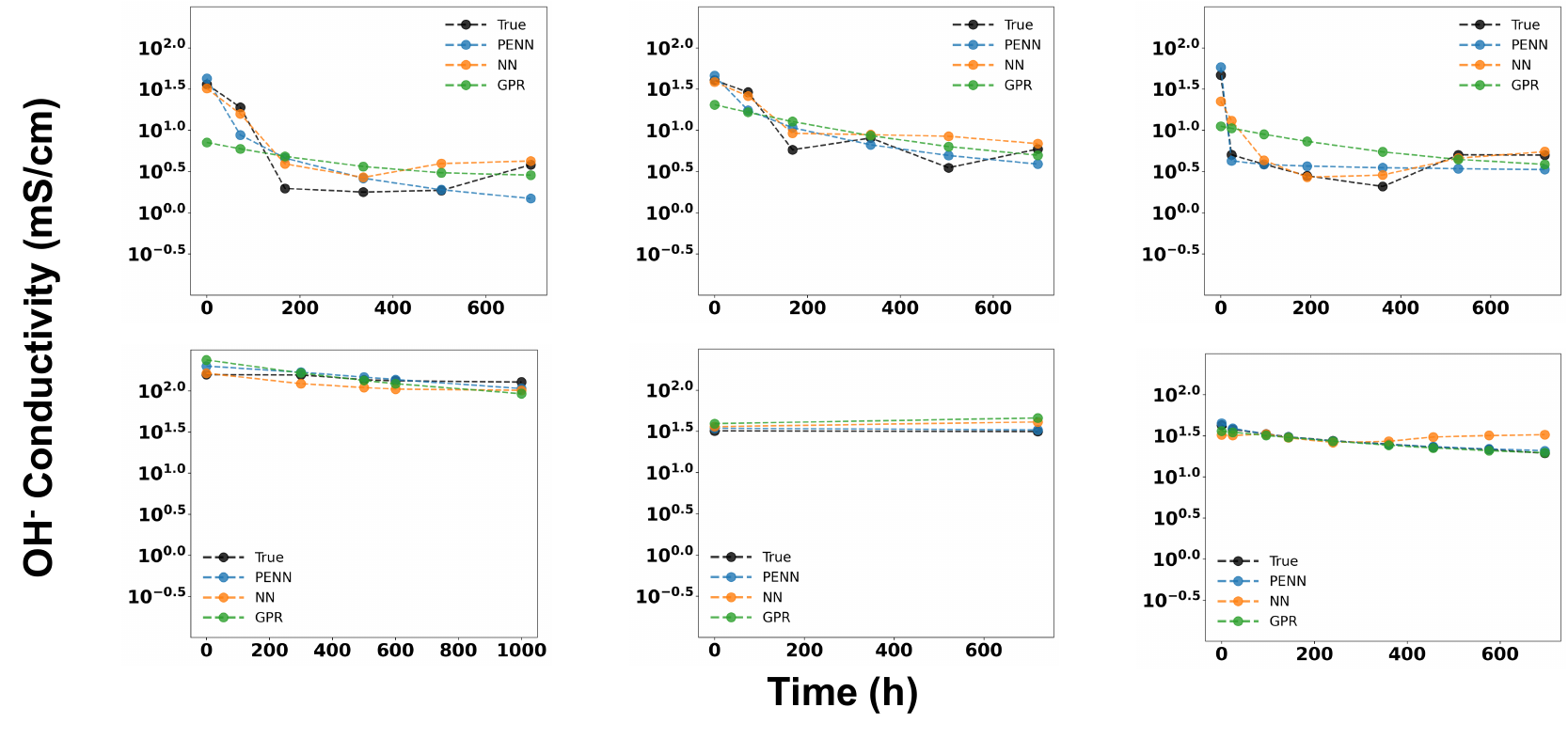}
    \caption{Representative degradation curves comparing PENN (blue), NN (orange) and GPR (green) predictions against experimental data (black) for six different AEM samples. Each model was trained on all available data. The top row depicts cases with more drastic degradation, while the bottom row depicts more moderate degradation profiles.}
    \label{fig:deg_pred_ex_train}
\end{figure}

\subsection*{Machine Learning Framework}
Three machine learning (ML) models were implemented to model time-dependent degradation in AEMs: Gaussian Process Regression (GPR), a classic Neural Network (NN) and a Physics-Enforced Neural Network (PENN). GPR and NN served as non-physics baselines, while PENN incorporated physical constraints into its architecture to capture degradation dynamics.

\subsubsection*{GPR and NN: Non-Physics Baselines}
GPR was implemented as a non-parametric Bayesian regression method capable of producing both point predictions and associated uncertainty estimates \cite{zou_machine_2021, phua_predicting_2023}. Hydroxide conductivity was predicted directly from a feature vector including polymer fingerprints, environmental variables, additive descriptors, and time as an explicit input feature. A composite kernel combining a radial basis function and white noise was employed to capture both smooth nonlinear relationships and experimental noise. GPR models were trained using Scikit-learn \cite{pedregosa_scikit-learn_2011} with five-fold cross-validation across five random seeds. Model performance was averaged over the folds and seeds to ensure statistical robustness.

NN was implemented as an additional baseline to see if a simple neural network approach could mitigate the issues with GPR, or if a physics-enforced architecture was necessary for this application. The NN was implemented as a fully-connected feedforward neural network model using PyTorch \cite{paszke_pytorch_2019}. The network comprised an input layer matching the dimension of the fingerprinted feature vector (chemical descriptors, time, etc), three hidden layers with nonlinear activation functions and dropout layers, and an output layer predicting the log of conductivity. Hyperparameter optimization was performed using Optuna \cite{akiba_optuna_2019}, which employed Bayesian optimization to explore:

\begin{itemize}
    \item learning rate (1$\times10^{-4}$–1$\times10^{-3}$),
    \item hidden layer sizes (Layer 1: 512–1024 units; Layer 2: 128–512 units; Layer 3: 64–128 units),
    \item dropout rates (0.1–0.5 for all layers).
\end{itemize}

The final model configuration corresponded to the hyperparameter set yielding the lowest training loss. Models were trained using the Adam optimizer with a learning rate scheduler that reduced the learning rate by a factor of 0.5 after 100 epochs without validation loss improvement, with a lower bound of 1$\times10^{-6}$. Training was terminated using an early stopping criterion after 200 epochs without improvement in validation loss. Loss was defined as the mean squared error between the predicted and true conductivity.

\subsubsection*{PENN: Physics-Enforced Neural Network}

The PENN model was implemented similarly to the NN model described above, with a few key differences. As illustrated in Figure \ref{fig:PENN_architecture}, instead of including time as an input feature and directly predicting conductivity, the architecture was built such that the final layer was a vector of length four, with each component of the final vector corresponding to one of the scalar parameters of the degradation profile. These parameters were then scaled to align the predicted values with the empirically observed ranges and finally plugged into Equation \ref{deg} along with each sample's recorded time value to get a conductivity prediction. Hyperparameter optimization was again performed using Optuna, which explored the same parameter space as the NN case but with the addition of the physics weight parameter $\omega$ (0.00–0.50, step size 0.01), which is discussed in the PENN Architecture Design section.

\begin{figure}[t]
\centering
\includegraphics[width=0.75\linewidth]{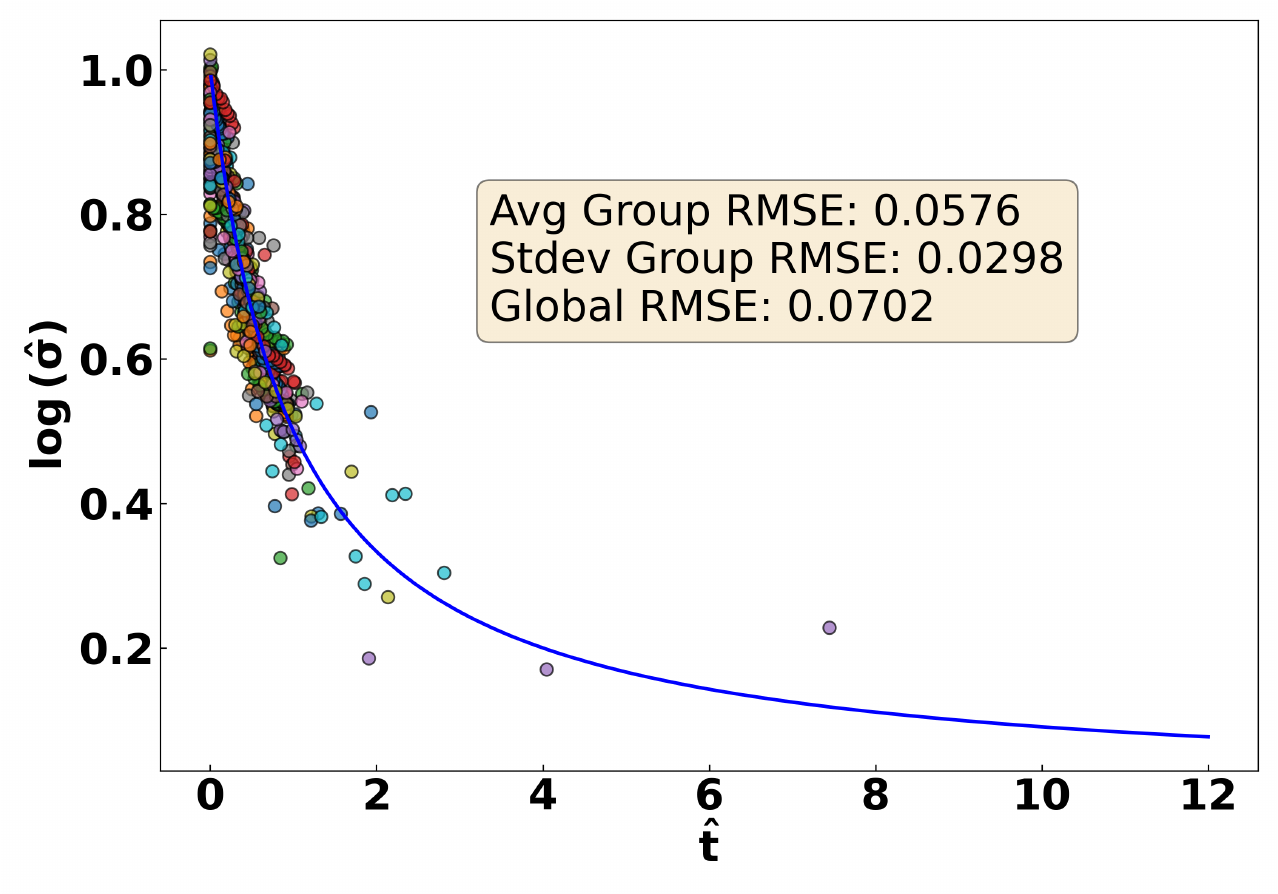}
\caption{Normalized degradation behavior across all AEM samples. The PENN-predicted degradation curves collapse onto a universal master curve defined by Equation~\ref{deg_normalized}. The blue line represents the idealized form \(y=\frac{1}{1+x}\). This agreement across chemistries and conditions reveals a shared empirical degradation mechanism and confirms the ability of the PENN to uncover universal trends.}
\label{fig:universal_fit}
\end{figure}

\section*{Results and Discussion}

\begin{figure}[t]
\centering
\includegraphics[width=0.75\linewidth]{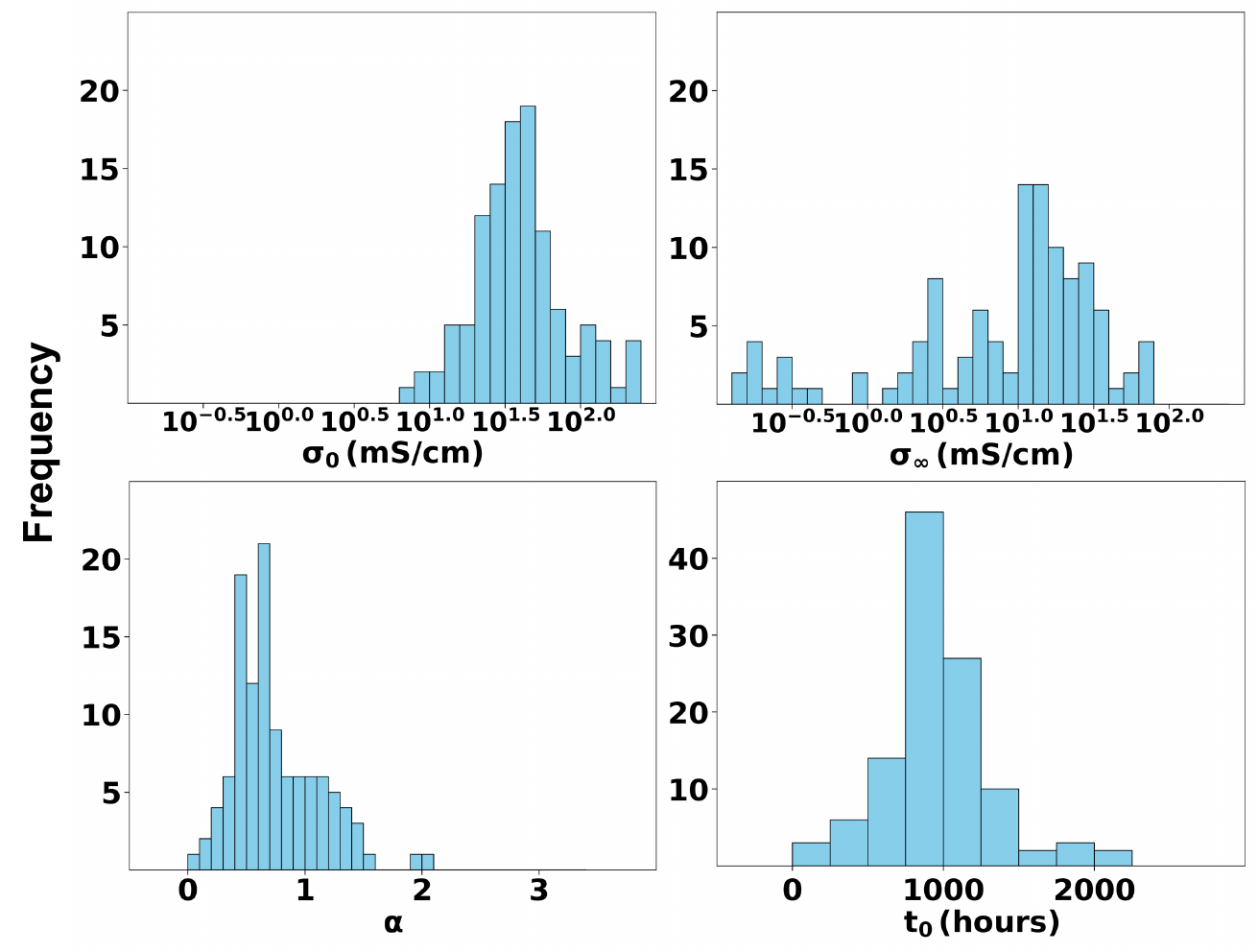}
\caption{Distribution of PENN-predicted degradation parameters across all AEM samples. Histograms show the learned values of \(\, \sigma_0\) (top left), \( \, \sigma_{\infty}\) (top right), \(\alpha\) (bottom left), and \(t_0\) (bottom right). These distributions reflect the variability in conductivity behavior across different chemistries and testing conditions, highlighting materials with sharper or more gradual degradation.}
\label{fig:param_distribution}
\end{figure}

\subsection*{Modeling Strategy for Time-Dependent Degradation}

A central challenge in modeling AEM degradation is how to incorporate time-dependent behavior without losing the identity of the underlying material. A simple approach is to treat time as an input feature—concatenated alongside polymer fingerprints, environmental conditions, and additive concentrations. However, this method implicitly assumes that an AEM sample observed at different time points corresponds to entirely different materials, ignoring the continuity of its degradation trajectory. To overcome this, we adopt a more physically meaningful strategy: we decouple time from the input vector and instead inject it directly into a custom loss function that evaluates model predictions against the full temporal degradation profile. This allows the model to learn the universal degradation dynamics from the data while preserving the chemical identity. We compare these approaches by benchmarking a GPR model and a classic NN which use time as an input feature against a PENN that learns degradation behavior by embedding time into the modeling process itself.

\subsection*{PENN Architecture Design}

To address the limitations of GPR and classic NN, we implemented a PENN framework that leverages a mechanistic model of conductivity degradation. Rather than predicting conductivity directly at each time point, the PENN is trained to predict the parameters of Equation \ref{deg}. The neural network learns to predict \(\sigma_0, \sigma_{\infty}, t_0,\) and \(\alpha\) for each sample given its feature vector. Time is not used as an input feature, but instead appears only in the loss function, where the predicted degradation curve is compared to the experimental conductivity time series. This formulation ensures that the model respects known physical behavior and enables accurate extrapolation beyond the training time window.

The loss function is defined as the mean squared error between predicted and true conductivity values after passing the four parameters and the time data for each sample in a particular batch through Equation \ref{deg}. Additional penalties are applied during training to enforce known physical constraints. Given a time-resolved sample, we can ascertain that the predicted \(\sigma_0\) should be greater than or equal to the first property measurement at \(t=0\). Similarly, the predicted \(\sigma_{\infty}\) should be less than or equal to the final property measurement, where t is the largest value. These constraints are intended to  modify the optimization landscape towards more physically relevant spaces. A small weighting parameter \(\omega\) is used to optimize the amount of emphasis placed on these additional constraints.

\subsection*{Comparison of GPR, NN and PENN Performance on Training Data}

We begin by comparing the performance of the GPR and NN baselines with PENN model on all time-resolved samples using all data for training. Figure~\ref{fig:parity_plots} presents parity plots for all three models, where PENN achieves an overall \(R^2\) of 0.987, slightly higher than NN's \(R^2\) 0.955 or GPR’s \(R^2\) of 0.951. Although this small numerical gap might suggest similar performance, a closer inspection of degradation forecasting profiles reveals significant differences. 

Figure~\ref{fig:deg_pred_ex_train} shows representative degradation profiles that highlight the advantage of PENN over GPR and NN in capturing the physics of AEM degradation. By learning to predict physical parameters associated with observed degradation trends, PENN’s physics-informed architecture captures both the sharp early transition and the subsequent leveling-off phase, resulting in more accurate overall predictions and a more reliable estimate of the time required to reach stabilization. In contrast, GPR’s reliance on a stationary kernel over-smooths the early rapid changes, causing it to underestimate the initial drop and introduce slight misalignment in long-term predictions, and NN models tend to overfit the training data, providing nonphysical predictions (increasing conductivity with time).

This ability to represent distinct degradation phases is critical for meaningful long-term forecasting. In real membranes, an initial period of rapid damage is often followed by a slower, stabilization-driven decay, and PENN’s mechanistic constraints allow the model to accurately capture this two-stage behavior. As a result, PENN produces predictions that are not only more accurate but also more faithful to the underlying physical processes driving membrane degradation. Importantly, as will be demonstrated in the forecasting section, the ability to distinguish between these degradation regimes could inform the design of next-generation AEMs—enabling targeted materials development for applications where a short burst of high power output is acceptable, as well as for scenarios demanding long-term, stable performance.

\begin{figure}[ht!]
\centering
\includegraphics[width=\linewidth]{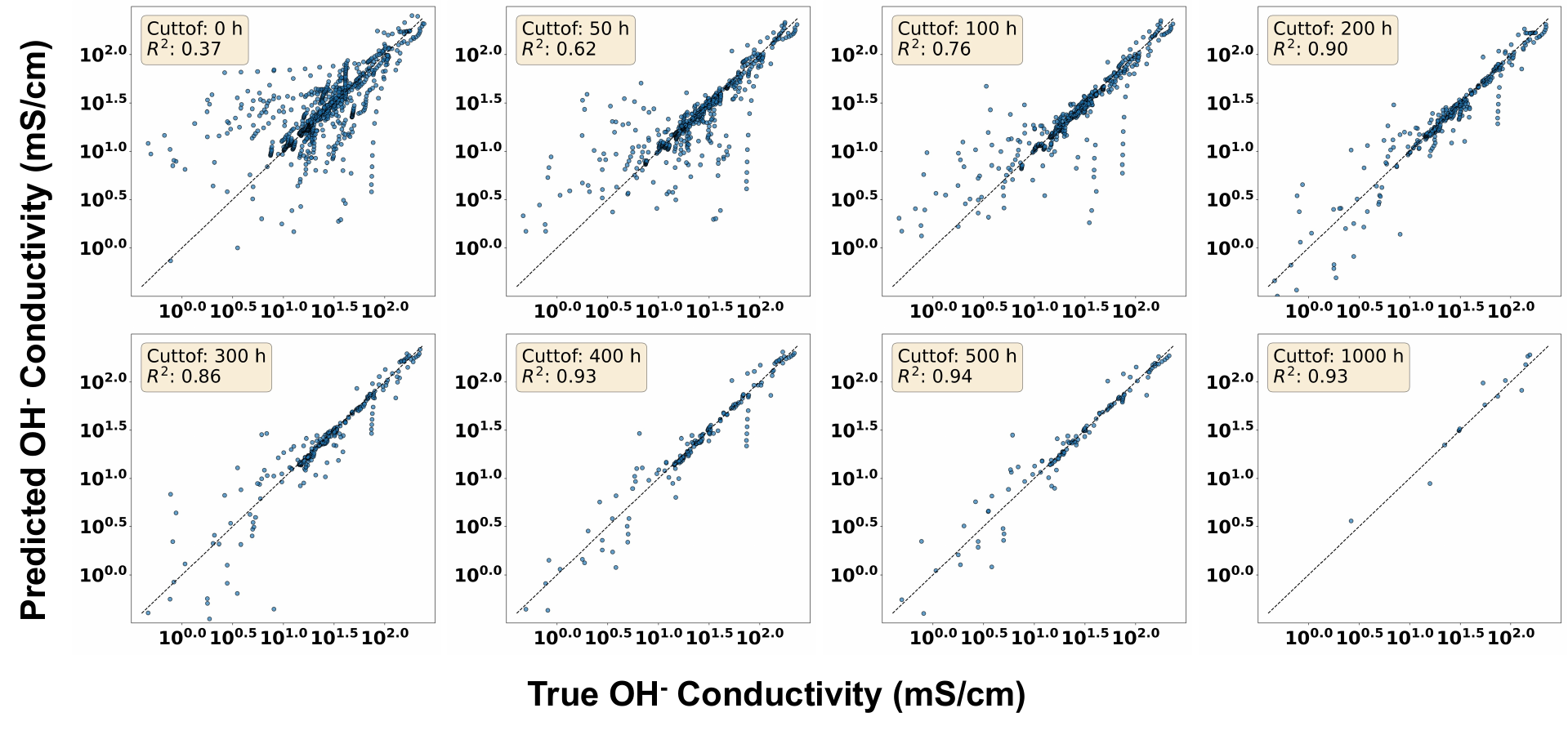}
\caption{Parity plots of predicted and true hydroxide conductivity for each cutoff value using the PENN models. Models were trained on all available data except the portion of each sample’s data beyond the designated cutoff time (0–1000 h); data from other samples beyond that cutoff remained available for training.}
\label{fig:forecasting_parity}
\end{figure}

\begin{figure}[ht!]
\centering
\includegraphics[width=0.6\linewidth]{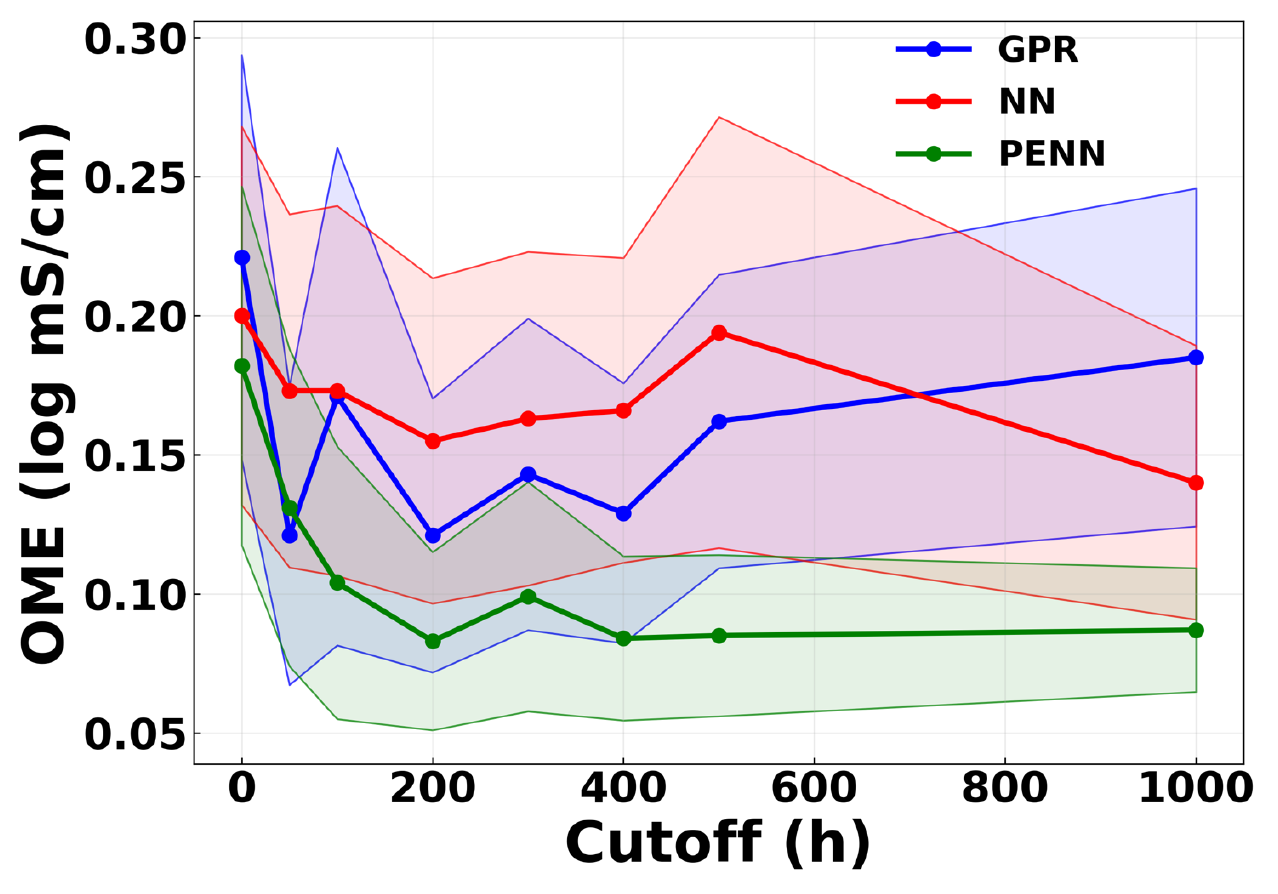}
\caption{Average order-of-magnitude error (OME) as a function of cutoff time for GPR, NN, and PENN models. Each data point represents the mean prediction error across all test samples withheld from training at a given cutoff. Shaded regions denote $\pm 0.25\sigma$ (one-quarter of the standard deviation) around the mean OME, illustrating variability across cutoff values and training algorithms.}
\label{fig:forecasting_ome}
\end{figure}

\subsection*{Emergence of a Universal Degradation Curve}

By applying the PENN model across the full dataset, we observe that the predicted degradation behavior of all samples collapses onto a single, normalized master curve when plotted using the rescaled variables defined in Equation~\ref{deg_normalized}. This result, shown in Figure~\ref{fig:universal_fit}, confirms our hypothesis that despite the chemical and environmental diversity in the dataset, degradation follows a shared empirical trajectory. This universal behavior reveals a powerful abstraction: conductivity decay in AEMs can be effectively parameterized using just four physically meaningful quantities. The ability to normalize this behavior across systems is crucial for guiding future design by establishing performance benchmarks and degradation archetypes.

We further analyze the distribution of predicted parameters \(\sigma_0\), \(\sigma_{\infty}\), \(t_0\), and \(\alpha\) across the dataset. The histograms shown in Figure~\ref{fig:param_distribution} highlight trends such as the clustering of \(\alpha\) between 1.5 and 3.0, the broader variation in \(t_0\), and the bimodal distribution of \(\sigma_{\infty}\), indicating differences in degradation kinetics between chemistries. These distributions offer insight into material design: a high \(\alpha\) value corresponds to sharper decay after an initial stable region, whereas longer \(t_0\) implies greater resistance to degradation. Such correlations can inform rational design strategies for more durable AEMs or those with high energy burst capabilities but long-term susceptibility to degradation.

In this scheme, models are trained on the entire dataset of hydroxide conductivity profiles and predictions are made for each time-resolved sample. Then, after predicting the four parameters for each sample, they are plugged into equation \ref{deg_normalized} to compare the fit of all of the time-resolved samples at once. 
The goal of this approach is to identify generalizable patterns that govern the degradation of anion exchange membranes across diverse chemistries, processing conditions, and environmental exposures. By training on the complete time series data for each polymer, the model learns to capture the underlying structure of conductivity decay across thousands of degradation trajectories. For the PENN model, this enables the identification of a normalized degradation manifold that describes how conductivity evolves as a function of scaled time, independent of specific chemical details. This universal trend is particularly useful for uncovering shared degradation mechanisms and benchmarking materials against common decay baselines.

\begin{figure}[!t]
    \centering
    \includegraphics[width=0.95\linewidth]{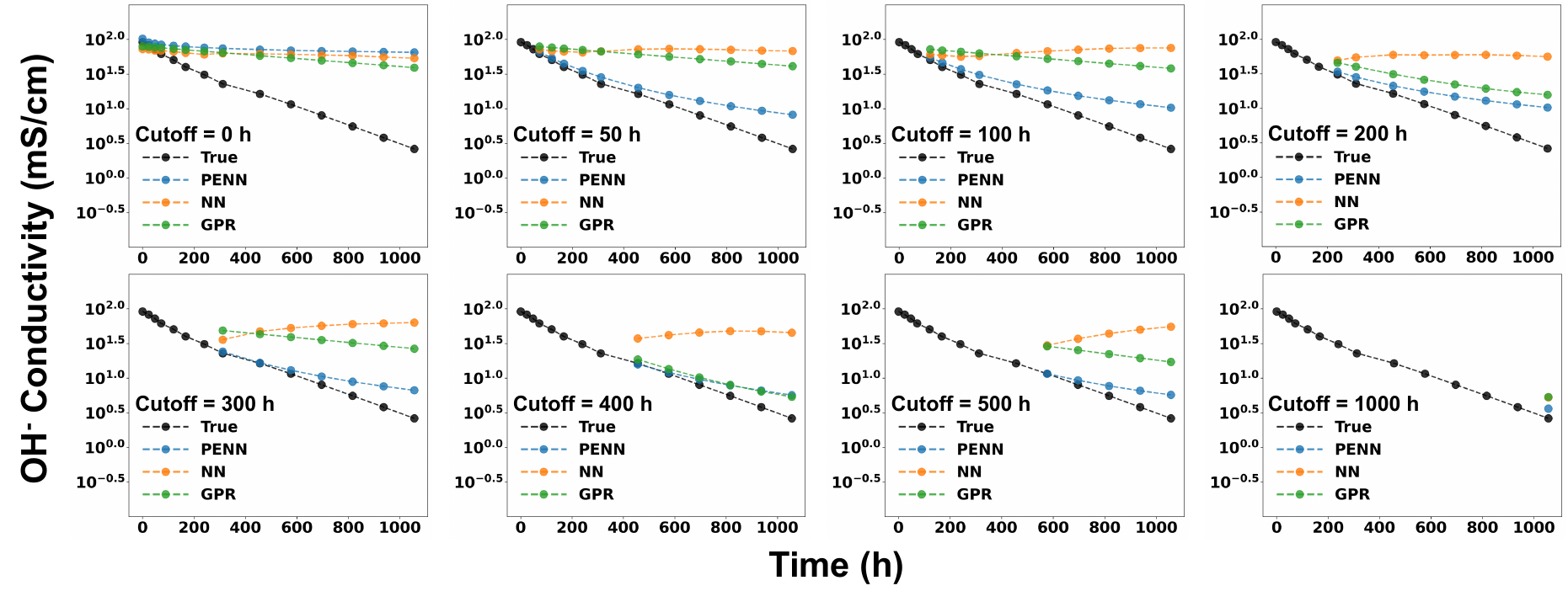}
    \caption{Representative degradation forecasting curves comparing PENN (blue), NN (orange) and GPR (green) predictions against experimental data (black) for a single AEM sample across a range of cutoff values (trained on data for all other samples plus the data up until the cutoff point, and predicted on data after the cutoff point). Predictions using the PENN models drastically improve with small amounts of data, and become more accurate with the inclusion of more data. NN models behave nonphysically (increasing conductivity predictions with time) and GPR models are unable to match the trend predicted by PENN.}
    \label{fig:deg_pred_ex}
\end{figure}

\subsection*{Forecasting Long-Term Degradation from Early-Time Data}

One of the key strengths of the PENN framework is its ability to forecast long-term degradation from short-term measurements. We implement a time-threshold validation strategy to assess the model's ability to forecast long-term degradation from limited early-time data. For each time-resolved AEM sample, we construct a model that is trained on the full dataset excluding that sample’s later-time measurements. Specifically, for a given sample, only data points prior to a selected time threshold are included, while all data from other samples are retained. We repeat this procedure for every sample and for a series of thresholds: 0 h (no data from this sample is included in the training set), 50 h, 100 h, 200 h, 300 h, 400 h, 500 h, and 1000 h. Figure~\ref{fig:forecasting_parity} shows PENN parity plots for the various time thresholds and indicates that with no data for a particular sample, degradation forecasting predictions are reasonable, and that even with as little as 200 h of data, PENN achieves accurate predictions of conductivity up to 10,000 h for most samples. The improvement in performance with increasing threshold demonstrates the value of early-time data while also quantifying the point at which degradation forecasting becomes reliable. This capability is critical for real-world applications, where prolonged testing is often infeasible. The results from this exercise (as highlighted in the average Order of Magnitude Error (OME) vs threshold plot in Figure ~\ref{fig:forecasting_ome}) indicate that after about 200 hours of performance data there is a significant diminishing returns to collecting longer-time data to predict longer time behavior, and that the PENN models significantly outperform GPR and NN in forecasting ability with limited data, making it an ideal approach for future AEM design schemes. As reflected by the shaded regions in Figure~\ref{fig:forecasting_ome}, which represent $\pm 0.25\sigma$ (one-quarter of the standard deviation) around the mean OME, the PENN model also exhibits lower variability across samples, indicating greater robustness and generalization capability.

An example degradation profile for one sample across each cutoff value for each model is shown in Figure \ref{fig:deg_pred_ex}. The results enforce the benefit of using physics-based modeling in long-term degradation performance as a cost-saving measure for materials design initiatives. This approach mimics realistic experimental constraints, where extended aging studies may be infeasible due to time, cost, or material limitations. By evaluating model performance at increasing time thresholds, we identify the earliest time point at which partial degradation data becomes predictive of long-term behavior. This analysis provides insight into the temporal data requirements for reliable forecasting and supports the design of efficient experimental protocols. Ultimately, this forecasting capability enables rapid, data-efficient screening of AEM candidates based not only on their initial properties but also on their projected durability.

\section*{Conclusions, Limitations, and Future Work}

We introduced a physics-enforced neural network (PENN) that couples polymer-genome fingerprints and environmental descriptors with a mechanistic degradation equation to model the time evolution of hydroxide conductivity in AEMs. Using a literature-curated dataset of time-resolved measurements, PENN (i) learns four interpretable parameters ($\sigma_0,\ \sigma_{\infty},\ t_0,\ \alpha$) that quantify initial performance, long-term limits, timescales, and decay shape; (ii) reveals a normalized universal degradation curve across diverse chemistries and conditions; and (iii) outperforms baseline NN and GPR models in forecasting long-term behavior from sparse early-time data. Practically, we find that \(\sim\)200~h of measurements often suffices to enable accurate extrapolation toward thousands of hours, reducing experimental burden while preserving physical fidelity. These capabilities position PENN as a data-efficient, interpretable framework for accelerated AEM screening and design based on both initial performance and projected lifetime.

While the PENN framework offers strong predictive performance and interpretability, several limitations should be acknowledged:

\begin{itemize}
    \item \textbf{Fixed Functional Form:} The degradation model assumes that conductivity decay universally follows Equation~\ref{deg}. While this empirically fits the data well, real-world degradation may involve multistage or non-sigmoidal dynamics in some chemistries.
    \item \textbf{Chemical Space Limitations:} The model generalizes well within the training distribution, but predictions for novel backbones or additives outside the dataset may suffer from extrapolation uncertainty.
    \item \textbf{Single-Property Focus:} This work focuses solely on hydroxide conductivity. However, mechanical degradation and dimensional stability are also critical to AEM lifetime. A future extension to multi-task PENNs could jointly model conductivity, swelling, and tensile degradation.
    \item \textbf{Experimental Validation:} Model predictions—especially forecasts beyond 1000 h—should be validated experimentally. PENN provides hypotheses for long-term behavior but should be used as a screening and guidance tool.
\end{itemize}


Overall, the PENN framework demonstrates superior robustness, physical consistency, and generalizability compared to traditional regression models and neural networks. By integrating domain-specific constraints and leveraging a parameterized degradation equation, it enables accurate modeling across diverse chemical and environmental conditions, identification of universal trends in AEM degradation, quantification of meaningful degradation parameters, and reliable forecasting from sparse experimental data. This modeling framework provides a foundation for accelerated screening of AEM candidates, allowing researchers to prioritize materials based not only on their initial performance but also their projected lifetime.

\section*{Acknowledgements}
This work was supported as part of the UNCAGE-ME, an Energy Frontier Research Center
funded by the U.S. Department of Energy, Office of Science, Basic Energy Sciences at the
Georgia Institute of Technology under award \# DE-SC0012577.

\bibliography{bib}

\end{document}